\documentclass[aps,prb,showpacs]{revtex4-1}
\usepackage{latexsym}
\usepackage{amsfonts}
\usepackage[english]{babel}
\usepackage[dvips]{epsfig}
  \usepackage{xcolor}[2007/01/21]
\usepackage[dvipdfm,bookmarks=true,bookmarksnumbered,bookmarksopen]{hyperref}
\hypersetup{pdfpagemode=FullScreen,colorlinks=true,
citecolor=green,unicode,bookmarksopenlevel=3,
pdftoolbar=true,pdfauthor={Helen Gomonay},pdftitle={On the theory
of competition between ferro- and antiferromagnetic domains in
multiferroics}}
\begin{document}
\title{On the theory of magnetization in multiferroics: competition between ferro- and antiferromagnetic domains}
\author{Helen V. Gomonay$^{1,2)}$, Ieugeniya G. Korniienko$^{1)}$, and Vadim M. Loktev$^{1,2)}$}
\affiliation{~$^{1)}$National Technical University of Ukraine
``KPI'', ave Peremogy, 37, 03056 Kyiv, Ukraine
\\ ~$^{2)}$ Bogolyubov Institute for Theoretical Physics NAS of Ukraine,\\ Metrologichna str. 14-b, 03680, Kyiv,
Ukraine\\
E-mail: {\rm Helen Gomonay} $\langle{\tt
malyshen@ukrpack.net}\rangle$}

\begin{abstract}
Many technological applications of multiferroics are based on their
ability to reconstruct the domain structure (DS) under the action of
small external fields. In the present paper we analyze the different
scenarios of the DS behavior in a multiferroic that shows
simultaneously ferro- and antiferromagnetic ordering on the
different systems of magnetic ions. We consider the way to control a
composition of the DS and
 macroscopic properties of the sample by an appropriate field
 treatment. We found out that sensitivity of the DS to the  external magnetic field and the magnetic susceptibility in a
 low-field region are determined mainly by the destressing effects (that have
 magnetoelastic origin). In a particular case of Sr$_{2}$Cu$_{3}$O$_{4}$Cl$_{2}$ crystal we anticipate the
 peculiarities of the elastic and magnetoelastic properties at
 $T\approx 100$~K.
\end{abstract}
\maketitle
\section{Introduction}\label{introduction}
During the last ten years a special attention is paid to the
materials in which magnetism coexists with the other types of
ordering, i.e., ferroelectric \cite{Fert:2007}, elastic
\cite{cruz:2006}, martensitic \cite{Chernenko:1995}. Solids that
show strong coupling between the different types of ordering are
often called multiferroics \cite{Ramesh:2007}. Growing interest to
multiferroics is based on the possibility to \emph{i}) control such
macroscopic properties of a sample as conductivity, magnetization,
elongation, with the suitable fields of different nature; \emph{ii})
manipulate the state of the magnetically (electrically, etc.) inert
materials (see, e.g., Refs. \onlinecite{Chu:2007,desousa-2008-3}).

One of the most technologically appealing property of multiferroics,
namely, sensitivity of their  macroscopic properties to the
influence of small external fields is due to formation and
reconstruction of the domain structure (DS) \cite{Fiebig:2002}. This
adaptivity, ability to change macroscopic parameters (such as a
shape, magnetization, electric polarization) in response to external
forces is related with the finite size and boundary of the sample.
While the physical mechanism of the DS formation is related with the
sample boundary, reconstruction and restructurization of the domains
under external fields depends upon the properties of the domain
walls. If a potential barrier for the domains wall formation is
high, switching between the different macroscopic states is sharp
and field dependence of macroscopic parameters reveals a hysteresis.
In the opposite case of low potential barrier, reconstruction of the
DS takes place through the nucleation and growth of new domains and
shows the features of liquid-like behavior: nonhysteretic
transitions between the different macroscopic states, shape
deformation, etc. The most interesting case on which we concentrate
our attention in the present paper lies in-between: in multiferroics
with the domains of different nature some types of the domains can
easily nucleate and show soft-like behavior while the others could
have high nucleation barrier and reveal themselves as solids.

The origin of the DS in the ``single'' ferroics, like ferromagnets
(FM) and ferroelectrics, is well established \cite{Kittel:1949} and
is attributed to the presence of long-range interactions between the
magnetic or electric dipoles localized on different sites. The
nonlocal character of the dipole forces ensures strong dependence of
the equilibrium DS on the shape of the sample. In many important
cases the DS of a single ferroic consists of the domains with
opposite (sometimes noncollinear) directions of polarization and can
be described thermodynamically on the basis of demagnetization
energy.

Antiferromagnets (AFM)  give an example of the more complicated
materials with usually pronounced coupling between the magnetic
order parameter and lattice strain\footnote{~In general, FM
materials also show magnetoelastic coupling. However, anisotropic
magnetostriction in AFM may have an exchange origin and thus may be
much larger than that in FM. Moreover, in contrast to AFM,
ferromagnetic domains with opposite direction of magnetization could
be easily distinguished by the magnetic field but show the same
magnetoelastic strain.}. The behavior of equilibrium domain
structure in AFM is very similar to the behavior of the DS in other
ferroics with some ``technical'' distinctions: \emph{i}) long-range
dipole-dipole interactions responsible for the formation of
equilibrium DS have a magnetoelastic origin and are described by the
destressing \cite{gomonay:174439} (in contrast to depolarization or
demagnetization) energy; \emph{ii}) DS consists of the domains with
different (nonparallel) orientations of AFM vectors.

Description of the DS in multiferroics seems to be much more
complicated problem mostly due to the fact that the domains of
different nature appear at different scales and thus form a
hierarchical structure. Good example of such a complexity is given
by FM martensites (see, e.g., Ref.~\onlinecite{Chernenko:1995}) that
are usually characterized by two independent order parameters
\cite{lvov:1998}, magnetization and spontaneous
strain\footnote{~Ferromagnets with the pronounced magnetostriction
may also show nontrivial DS. In contrast to FM martensites, the DS
in such materials consists of the domains that can be appropriately
described by one (and only one) order parameter, magnetization or
strain tensor component.}. Cross-correlations between the magnetic
and structural order parameters open a way to control a martensitic
DS and, as a consequence, macroscopic deformation of a sample, with
the external magnetic field (so-called giant magnetostriction
\cite{Kokorin:1996(1)}).

\begin{figure}[htbp]
 \includegraphics[width=0.7\columnwidth]{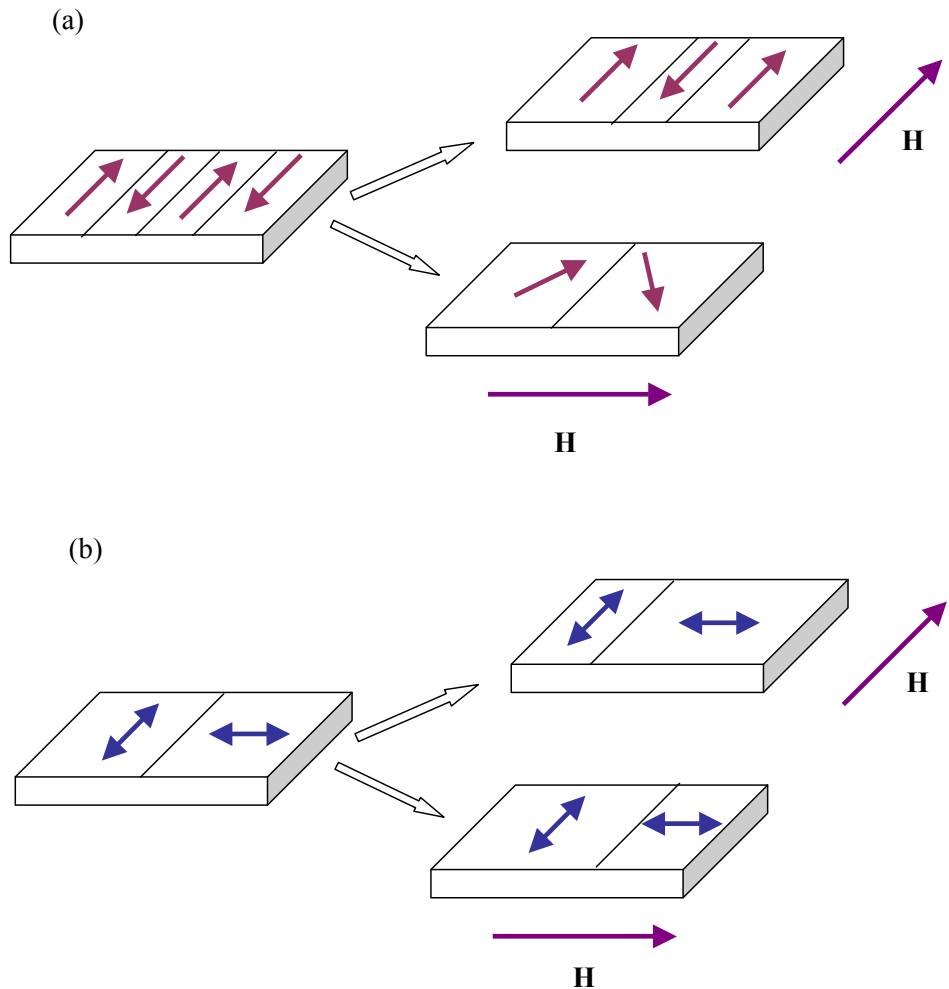}
  \caption{(Color online) Behavior of the FM (a) and AFM (b)
domain structure in the external magnetic field $\mathbf{H}$. In the
absence of field both types of domains (shown by arrows) are equally
represented. (a) FM domains have opposite direction of magnetization
vector. The magnetic field applied parallel to an easy axis (upper
panel) removes degeneracy of the domains. As a result, fraction of
the favorable domain increases. If $\mathbf{H}$ is perpendicular to
the easy axis, domains of both types are equivalent, the domain
fraction does not change, magnetic field induces a tilt of the
magnetizations (lower panel). (b) AFM domains have different
(perpendicular) orientations of AFM vectors. Degeneracy of the
domains is removed for any of two mutually perpendicular
orientations of the magnetic field. \label{fig_fm_vs_afm_H}}
\end{figure}

Another example of multiferroic behavior is given by some of the
high-temperature superconducting systems (like
Sr$_{2}$Cu$_{3}$O$_{4}$Cl$_{2}$ or Ba$_{2}$Cu$_{3}$O$_{4}$Cl$_{2}$)
that show simultaneously FM and AFM ordering on the different
systems of copper ions. In contrast to FM martensites, the DS in
these crystals is not hierarchical.  Each type of domains is
characterized by two independent (FM and AFM) order parameters.
Though macroscopic state of both ferro- and antiferromagnets could
be controlled by the same, magnetic, field, the responses of the FM
and AFM domain structures are different, as illustrated in
Fig.~\ref{fig_fm_vs_afm_H}.
The domain structure of FMs reconfigures in the magnetic field which
is parallel to an easy axis and does not change if the magnetic
field is perpendicular to this axis. Macroscopic magnetization of
the sample (and hence, macroscopic susceptibility) is inversely
proportional to the appropriate component of demagnetization tensor.
In contrast, in antiferromagnetic crystals the DS reconfigures for
both mutually perpendicular orientations of the magnetic field.
Macroscopic magnetization depends upon the components of destressing
tensor that have a magnetoelastic origin. So, a material that bears
simultaneously the features of FM and AFM can show some new type of
behavior in the external magnetic field governed by competition
between the demagnetizing and destressing effects. 

In the present paper we study an equilibrium DS of multiferroic
Sr$_{2}$Cu$_{3}$O$_{4}$Cl$_{2}$  with the FM and AFM order
parameters. In the framework of phenomenological approach we analyze
the possible magnetization curves that could be obtained for the
samples of different shape and different field treatment. On the
basis of the developed model we make an attempt to
 interpret the unusual behavior of macroscopic
magnetization observed in the experiments of Parks et al
\cite{Parks:2001} and predict a peculiarity of the elastic
properties of Sr$_{2}$Cu$_{3}$O$_{4}$Cl$_{2}$ at the temperature
$T\approx100$~K.

\section{Model}
\begin{figure}[htbp]
  \includegraphics[width=0.5\columnwidth]{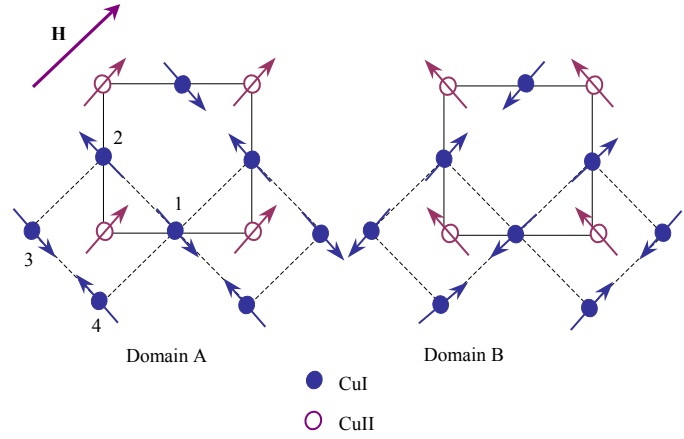}
  \caption{(Color online) Magnetic structure of Cu$_{3}$O$_{4}$ layer in two
  different
  configurations (domains). Magnetic field is parallel to $\langle
  110\rangle$. Two types of magnetic ions are represented with the filled and hollow circles. FM ordered moments of CuII could be (a) parallel (domain A) or (b) perpendicular (domain B) to the applied magnetic
  field. Small canting of the CuI spins induced by the external magnetic
  field is not shown.
\label{fig_crystal_dom}}
\end{figure}

  The crystal structure of high-temperature superconducting cuprates Sr$_{2}$Cu$_{3}$O$_{4}$Cl$_{2}$ and
Ba$_{2}$Cu$_{3}$O$_{4}$Cl$_{2}$ consists of Cu$_{3}$O$_{4}$ planes
separated by spacer layers of SrCl or BaCl \cite{Noro:1994,
Kim:2001, Parks:2001}. Two types of magnetic ions, CuI and CuII
(see Fig.~\ref{fig_crystal_dom}) form two interpenetrating square
lattices within Cu$_{3}$O$_{4}$ planes.
Within the temperature interval $T_{II}=40$~K$\le T \le T_{I}=380$~K
the ions of the first type (CuI) are AFM ordered while the ions of
the second type (CuII) bear small but nonzero FM
moment\footnote{According to Ref.\onlinecite{Kastner:1999}, the FM
moments at CuII ions result from the anisotropic ``pseudodipolar''
interactions between CuI and CuII.}. According to the experiments
\onlinecite{Kastner:1999},  mutual orientation of CuI and CuII
moments depends upon the direction of the external magnetic field
and can be either perpendicular or parallel. Thus, the magnetic
structure consists of two weakly coupled subsystems, namely, an AFM,
localized on CuI ions, and  a FM one, localized on CuII ions. The FM
subsystem is unambiguously described by the magnetization vector
$\mathbf{M}_{\rm F}$  and the AFM subsystem is described by two
vectors: AFM vector
$\mathbf{L}=(\mathbf{S}_1-\mathbf{S}_2+\mathbf{S}_3-\mathbf{S}_4)/4$
and ferromagnetic vector $\mathbf{M}=\sum_j\mathbf{S}_j/4$
(numeration of CuI sites is shown in Fig.~\ref{fig_crystal_dom}).

\begin{figure}[htbp]
  \includegraphics[width=0.5\columnwidth]{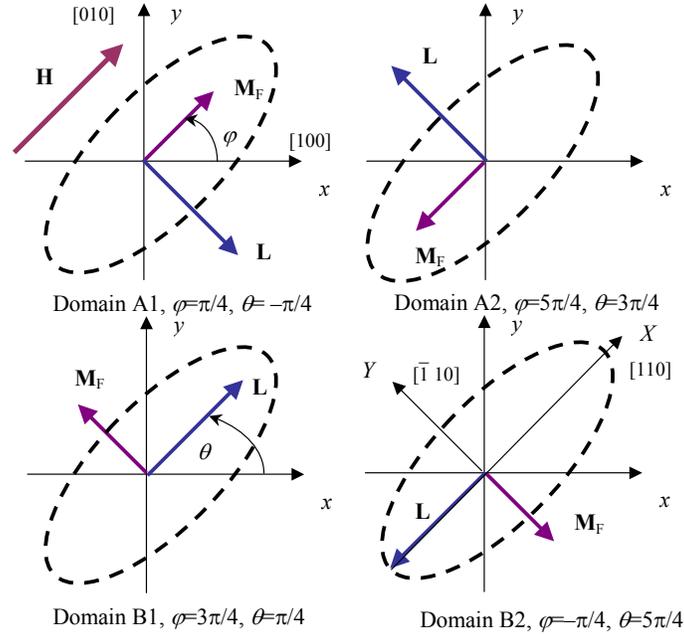}
  \caption{(Color online) Four types of magnetic domains. Axes $x$ and $y$ are parallel to $\langle 100\rangle$ crystal directions.
  The external magnetic field $\mathbf{H}\|[110]$ (if any). Types A
  and B have different orientations of AFM vector, types 1 and 2
  correspond to opposite directions of FM vector $\mathbf{M}_F$.
Ellipse (dash line) images the supposed shape of the sample and its
orientation (axes $X$, $Y$) with respect to crystal axes.
\label{fig_domain_types}}
\end{figure}

In the absence of the external field the FM moments at CuII sites
are oriented along $\langle 110\rangle$ crystal directions
perpendicular to the staggered magnetizations of AFM subsystem, as
shown in Fig.~\ref{fig_crystal_dom}. Due to tetragonal symmetry of
the crystal (space group $I4/mmm$) an equilibrium magnetic structure
can be realized in four types of equivalent domains as shown in
Figs.~\ref{fig_crystal_dom} and \ref{fig_domain_types}. Domains of
type A and B could be thought of as AFM domains because they
correspond to different orientations of $\mathbf{L}$ vector and thus
are sensitive to orientation of the magnetic field $\mathbf{H}$ with
respect to the crystal axes (see Fig.~\ref{fig_fm_vs_afm_H}). Types
A1 and A2 (and, correspondingly, B1 and B2) are FM domains, they
have an opposite direction of $\mathbf{M}_{\rm F}$ vector and could
be removed from the sample by $\mathbf{H}\|\mathbf{M}_{\rm F}$.

Phenomenological description of the DS is based on the analysis of
the free energy potential $\Phi$ of the sample. We take into account
three constituents of $\Phi$: magneticб $\Phi_{\rm{mag}}$, stray
(demagnetizing),  $\Phi_{\rm{stray}}$ and destressing,
$\Phi_{\rm{dest}}$, energy:
\begin{equation}
\label{free_energy_1} \Phi = \Phi_{\rm{mag}} + \Phi_{\rm{stray}} +
\Phi_{\rm{dest}}.
\end{equation}
Magnetic energy of Sr$_{2}$Cu$_{3}$O$_{4}$Cl$_{2}$ crystal in mean
field approximation is well established
\cite{Chou:1997,Kastner:1999, Kim:2001} and can be written as
follows:
\begin{eqnarray}\label{magnetic_energy_1}
  \Phi_{\rm{mag}}&=&\int_V dV\left\{\frac{4}{M_0^2}\left[J_0\left(\mathbf{M}^2-\mathbf{L}^2\right)+J_{\rm{av}}\mathbf{M}\cdot\mathbf{M}_{\rm F}+J_{\rm{pd}}\mathbf{M}_{\rm F}\hat
  \sigma_z\mathbf{L}+K_{\perp}L_z^2\right]\right.\nonumber\\
&-&\left.\frac{8}{M_0^4}K_{\|}L_x^2L_y^2-\mathbf{H}\cdot\mathbf{M}_{\rm
F }-2\mathbf{H}\cdot\mathbf{M}\right\}.
\end{eqnarray}
Here $V$ is the sample volume, $M_0$ is CuI sublattice
magnetization, orthogonal axes $x$ and $y$ are parallel to the
crystal directions [100] and [010], respectively (see
Fig.~\ref{fig_domain_types}). $\hat\sigma_z$ is the Pauli matrix.
The meaning and values of phenomenological constants are given in
Table~\ref{table_data}. In the last column of this Table all the
constants are converted to Oe by division by sublattice
magnetization $M_0=27.4$~Gs (that corresponds to spin $s=1/2$ per
CuI site).

\begin{table*}\caption{Parameters used in the free energy
[Eq.~(\ref{magnetic_energy_1})]. The second column gives the raw
data (in meV) as taken from Refs.\onlinecite{Chou:1997,Kastner:1999,
Parks:2001}, in the last column the same values are given in Oe. }
\label{table_data}
\begin{ruledtabular} \begin{tabular}{lccc}
 Parameter & Meaning
  & Value in meV
  & Value in Oe
      \\
\hline $J_0$& CuI-–CuI superexchange (in-plane)&130&1.02$\cdot
10^7$\\ \hline $J_{\rm av}$& isotropic pseudodipolar interaction
&-12&-9.4$\cdot 10^5$\\ \hline $J_{\rm pd}$& anisotropic
pseudodipolar interaction &-0.027&-2.1$\cdot 10^3$\\\hline
$K_{\perp}$& out-of-plane anisotropy\cite{Kim:2001} &0.068&$5.3\cdot
10^3$\\ \hline $K_{\|}$& in-plane anisotropy &10$\cdot
10^{-6}$&7.8$\cdot 10^{-2}$
\end{tabular}
\end{ruledtabular}
\end{table*}
Contributions $\Phi_{\rm{stray}}$ and $\Phi_{\rm{dest}}$ in
Exp.~(\ref{free_energy_1}) arise from the long-range dipole-dipole
interactions of the magnetic and magnetoelastic nature,
correspondingly, and depend upon the sample shape. We consider a
thin (thickness $c$) pillar with an elliptic crossection whose
principal axes $X$ and $Y$ are parallel to $\langle 110\rangle$
directions within the Cu$_3$O$_4$ layers (see
Fig.~\ref{fig_domain_types}). In this case
\begin{equation}\label{demagnetiz}
\Phi_{\rm{stray}}=\frac{V}{2}\left[N^{\rm{dm}}_a\langle
M_{\rm{F}X}+2M_X\rangle^2+N^{\rm{dm}}_b\langle
M_{\rm{F}Y}+2M_Y\rangle^2\right],
\end{equation}
where the brackets $\langle\ldots\rangle$ mean averaging over the
sample volume. The components of demagnetization tensor
$N^{\rm{dm}}_{a,b}$ are calculated in a standard way
\cite{Bar:1968E}:
\begin{equation}\label{demag_tensor}
 N^{\rm{dm}}_a=\frac{4\pi c}{a\sqrt{1-k^2}}\int_0^{\pi/2}\frac{\sin^2\phi d\phi}{\sqrt{1-k^2\sin^2\phi}};\quad N^{\rm{dm}}_b=\frac{4\pi c\sqrt{1-k^2}}{a}\int_0^{\pi/2}\frac{\cos^2\phi d\phi}{\sqrt{1-k^2\sin^2\phi}}.
\end{equation}
Here $a\ge b(\gg c)$ are the ellipse's semiaxes (parallel to $X$
and $Y$ axes) and the parameter $k^2=1-b^2/a^2$ depends upon an
aspect ratio $b/a$ of the sample.

The destressing energy can be written in an analogous form
\cite{gomonay:174439}
\begin{eqnarray}\label{destress_cylinder_2}
  \Phi^{\rm dest}&=&V\left\{N^{\rm des}_{\rm is}[\langle
  L_Y^2-L_X^2\rangle^2
+4\langle
L_XL_Y\rangle^2]\right.\nonumber\\
  &+&\left.N^{\rm des}_{\rm 2an} \langle
L_X^2-L_Y^2\rangle-N^{\rm des}_{\rm 4an}[\langle
  L_X^2-L_Y^2\rangle^2-4\langle
L_XL_Y\rangle^2]\right\}.
\end{eqnarray}
An explicit form of the destressing constants $N^{\rm des}$ depends
upon the elastic and magnetoelastic properties of the crystal which
we assume to be isotropic (that means, in particular, the following
relation between the elastic modula: $c_{11}-c_{12}=2c_{44}$). Then,
\begin{equation}\label{destress_constants_thin_film}
N^{\rm des}_{\rm is}=\frac{\lambda^2(3-4\nu)}{16c_{44}(1-\nu)},\quad
N^{\rm
des}_{2}=\frac{c}{b}\cdot\frac{[\lambda^2(2-3\nu)+\lambda_v\lambda]J_2(k)}{8c_{44}(1-\nu)},\quad
N^{\rm des}_{\rm
4an}=\frac{c}{b}\cdot\frac{\lambda^2J_4(k)}{3c_{44}(1-\nu)},
\end{equation}
where $\lambda$ and $\lambda_v$ are magnetoelastic
constants,
 $\nu=c_{12}/(c_{11}+c_{12})$ is the
Poisson ratio and we have introduced the dimensionless shape-factors
$J_{2,4}(k)$ as follows \cite{gomonay:174439}
\begin{eqnarray}\label{constants}
  J_2(k)&=&\int_0^{\pi/2}\frac{(\sin^2\phi+\cos2\phi/k^2)d\phi}{\sqrt{1-k^2\sin^2\phi}},\nonumber\\
J_4(k)&=&
\int_0^{\pi/2}\frac{(1-8\cos2\phi-k^2\sin^2\phi+8\cos2\phi/k^2)d\phi}{\sqrt{1-k^2\sin^2\phi}}.
\end{eqnarray}
In Eqs.(\ref{demagnetiz}) and (\ref{destress_cylinder_2}) we have
omitted $Z$($\|z\|[001]$)-components of the demagnetizing and
destressing tensors as inessential for further consideration.

Expressions (\ref{magnetic_energy_1}), (\ref{demagnetiz}), and
(\ref{destress_cylinder_2}) could be substantially simplified if one
takes into account that: \emph{i}) far below the N{\'e}el
temperature the values of sublattice magnetizations $M_0$ and
$M_{\rm F}$ are saturated  and constant; as a result \emph{ii})
$\mathbf{L}\perp\mathbf{M}$ and $\mathbf{L}^2+\mathbf{M}^2=M_0^2$
(normalization conditions); \emph{iii}) if the magnetic field is
much smaller than the spin-flip field, $H\ll J_0/M_0$ and coupling
between the FM and AFM subsystems is much smaller than AFM exchange,
$J_{\rm av}M_F\ll J_0M_0$, the magnetization induced in AFM
subsystem is small, $M\ll M_0$, and vector $\mathbf{M}$ can be
excluded from Eq.~(\ref{magnetic_energy_1}) as follows
\cite{Kosevich:1983}:
\begin{equation}\label{magnetization_AFM}
  \mathbf{M}=\frac{1}{8J_0}\left[\mathbf{L}\times\left[\left(\mathbf{H}-2\frac{J_{\rm av}}{M_0^2}\mathbf{M}_{\rm
  F}\right)\right]\right];
\end{equation}
\emph{iv}) if out-of-plane anisotropy is strong enough,
$K_{\perp}\gg K_{\|}$ (see Table~\ref{table_data}), all the
magnetic vectors lie within $xy$ (and, equivalently, $XY$) plane
and could be described with the only angle variable, as shown in
Fig.~\ref{fig_domain_types}:
\begin{equation}\label{angles_parametrization}
  L_x=M_0\cos\theta,\,L_y=M_0\sin\theta;\,M_{{\rm
  F}x}=m_{\rm F}M_0\cos\varphi,\,M_{{\rm
  F}y}=m_{\rm F}M_0\sin\varphi.
\end{equation}
Here $m_{\rm F}$(=$10^{-3}$ for Sr$_{2}$Cu$_{3}$O$_{4}$Cl$_{2}$
\cite{Kastner:1999}) is a dimensionless constant that represents
the ratio between the spin moments localized on CuII and CuI
sites.

With account of the relations (\ref{magnetization_AFM}) and
(\ref{angles_parametrization}) the specific potential $\phi\equiv
\Phi/V$ (see Exp.~(\ref{free_energy_1}) takes the following form
\begin{eqnarray}\label{magnetic_energy_2}
  \phi&=&4J_{\rm{pd}}m_{\rm F}\langle \cos(\theta+\varphi)\rangle+K_{\|}\langle \cos4\theta\rangle
  -\frac{J^2_{\rm{av}}}{8J_0}m^2_{\rm
  F}\langle\cos2(\theta-\varphi)\rangle\nonumber\\
  &-&m_{\rm F}H\left[\left(1-\frac{J_{\rm{av}}}{8J_0}\right)\langle\cos(\varphi-\psi)\rangle+\frac{J_{\rm{av}}}{8J_0}\langle\cos(2\theta-\psi-\varphi)\rangle \right]
  +\frac{H^2}{32J_0}\langle\cos2(\theta-\psi)\rangle
  \nonumber\\
&-&N^{\rm des}_{\rm 2an}\langle
\cos2(\theta-\psi)\rangle+\frac{1}{2}M_0m_{\rm
F}^2\left[N^{\rm{dm}}_a\langle\cos(\varphi-\psi)\rangle^2+N^{\rm{dm}}_b\langle\sin(\varphi-\psi)\rangle^2\right]\\
&+&N^{\rm des}\langle\cos2(\theta-\psi)\rangle^2+\Delta N^{\rm
des}\langle\sin2(\theta-\psi)\rangle^2.\nonumber
\end{eqnarray}

where  $\psi$ is an angle between the magnetic field and $x$-axis, $
N^{\rm des}\equiv N^{\rm des}_{\rm is}+N^{\rm des}_{\rm 4an}$,
$\Delta N^{\rm des}\equiv 4\left(N^{\rm des}_{\rm is}-N^{\rm
des}_{\rm 4an}\right)$, and we assume that the field is parallel to
one of the principal axes of the sample (this corresponds to the
experimental situation that will be discussed below). Here and for
the rest of the paper we use the values in Oe (see the last column
of Table~\ref{table_data}) instead of energy units (say,
$\phi\rightarrow \phi/M_0$, etc.).

Let us consider the case when the magnetic field is parallel to one
of the easy axes, $\mathbf{H}\|[110]$, so, $\psi=\pi/4$. In an
infinite sample (all the components of tensors $N^{\rm{dm}},N^{\rm
des}$ are equal to zero) minimization of $\phi$ with respect to
magnetic variables $\theta$ and $\varphi$ gives rise to the four
solutions labeled as  A1,2 and B1,2 (see
Fig.~\ref{fig_domain_types}). Equilibrium values at $H=0$ are
\begin{eqnarray}\label{equilibrium_domains}
&& {\textrm{state  A1}}: \qquad \theta_{\rm A1}=-\pi/4, \qquad
\varphi_{\rm A1}=\phantom{3}\pi/4; \nonumber\\
 && {\textrm{state  B1}}: \qquad \theta_{\rm B1}=\phantom{3}\pi/4, \qquad \varphi_{\rm
 B1}=3\pi/4;\\
  && {\textrm{state  A2}}: \qquad \theta_{\rm A2}=3\pi/4, \qquad \varphi_{\rm
  A2}=5\pi/4;\nonumber\\
&& {\textrm{state  B2}}: \qquad \theta_{\rm B2}=5\pi/4, \qquad
\varphi_{\rm B2}=-\pi/4.\nonumber
\end{eqnarray}

 It should be stressed that in contrast to pure AFMs
the configurations with $(\mathbf{M}_F, \mathbf{L})$ and
$(\mathbf{M}_F, -\mathbf{L})$ are inequivalent, due to anisotropic
pseudodipolar interactions (described by the constant $J_{\rm
pd}$).
\begin{figure}[htbp]
  \includegraphics[width=0.7\columnwidth]{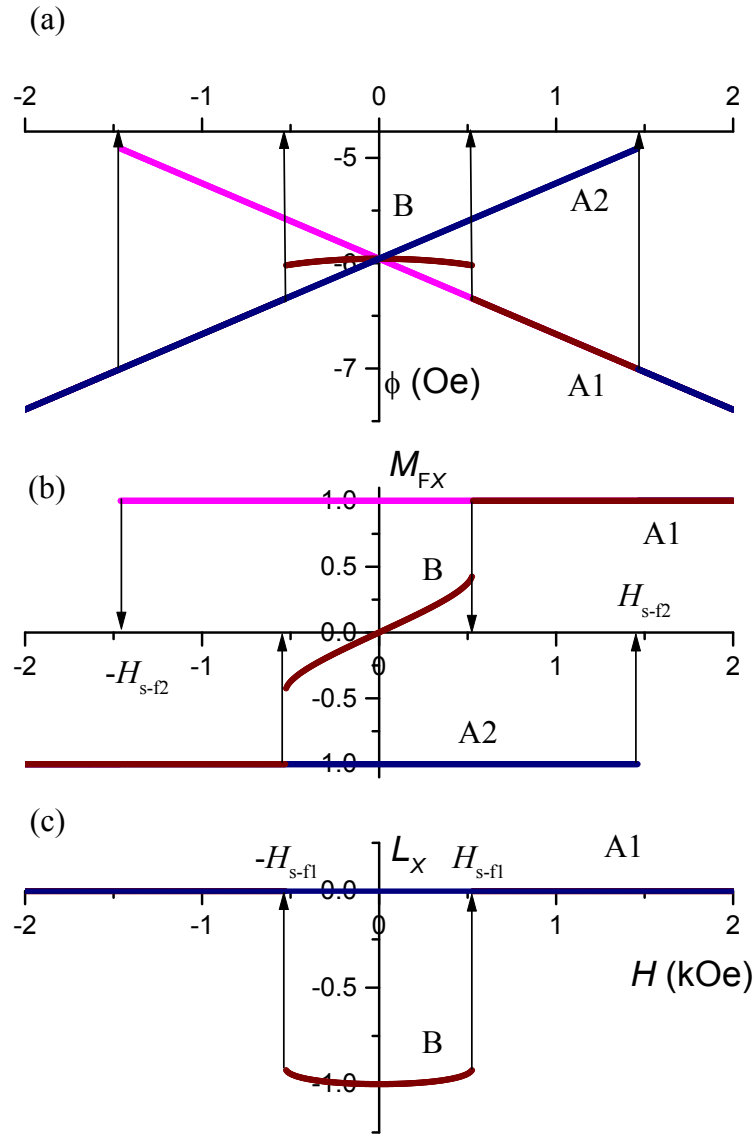}
  \caption{(Color online) Stability ranges of homogeneous
  configurations shown in Fig.~\ref{fig_domain_types} in the
  external magnetic field $\mathbf{H}\|[110]$. (a) Specific energy (in Oe) of
  equilibrium homogeneous state \emph{vs} $H$. (b), (c) Normalized
  projections of FM and AFM moments on the field direction. Field
  induces rather noticeable rotation of $\mathbf{M}_{\rm F}$
  vector toward the field direction (b) and slight tilt of
  $\mathbf{L}$ vector (c). Loss of stability takes place at
 the critical values $H=H_{\rm s-f1,2}$, as shown with
  arrows.\label{fig_homog_vs_field}}
\end{figure}

Fig.~\ref{fig_homog_vs_field} illustrates the field-induced
variation of equilibrium magnetic configurations (represented by
$X$-projections of $\mathbf{M}_{\rm F}$ and $\mathbf{L}$ vectors)
obtained from the numerical minimization of the
Exp.~(\ref{magnetic_energy_2}) using the data from
Table~\ref{table_data}. It is clearly seen that within the interval
$|H|\le H_{\rm s-f1}=525$~Oe there exist all four states A1,2 and
B1,2. The magnetic field removes degeneracy between the states A1,
A2 and B\footnote{~States B1 and B2 are equivalent in the field
parallel to [110] direction.}, as can be seen from
Fig.~\ref{fig_homog_vs_field}a. In particular, when $H\ge 0$, the
specific energies  $\phi_j\equiv \phi(\theta_j, \varphi_j)$ of
equilibrium states are related as follows: $\phi_{\rm A1}<\phi_{\rm
B}<\phi_{\rm A2}$. So, in some cases (discussed below) variation of
the external field may induce formation of the AFM (B) instead of
the FM (A2) domain. Orientations of $\mathbf{M}_{\rm F}$ and
$\mathbf{L}$ vectors in the A states are not influenced by the
field, while in the B states both vectors are slightly tilted (see
Fig.~\ref{fig_homog_vs_field}b,c). Rotation of AFM vector from the
field direction in the state B (where $\mathbf{H}\|\mathbf{L}$) is a
peculiar feature of the FM+AFM multiferroic caused by pseudodipolar
interactions between CuI and CuII ions. In the pure antiferromagnets
an AFM vector $\mathbf{L}$ keeps parallel (with respect to
$\mathbf{H}$) orientation  up to the field of spin-flop transition.

The first critical field $H_{\rm s-f1}\propto \sqrt{J_0 K_{\|}}$
corresponds to a step-like (spin-flop) transition
B1,2$\rightarrow$A1. In the interval $H_{\rm s-f1}<|H|<H_{\rm
s-f2}=1465$~Oe the potential $\Phi$ has only two minima that
correspond to the states A1 and A2. The second critical field
$H_{\rm s-f2}$ corresponds to 180$^\circ$ switching of
$\mathbf{M}_{\rm F}$ vector. Its value depends on the effective
anisotropy that originates from the pseudodipolar coupling
(corresponding constants $J_{\rm av}, J_{\rm pd}$) and in-plane
anisotropy $K_\|$ and can be calculated only numerically. Above
$H\ge H_{\rm s-f2}$ the sample is in a single domain state (A1).

\section{Equilibrium domain structure and magnetization curves}
On the large scales (much greater than the characteristic scale of
the magnetic inhomogeneity, i.e., domain wall thickness) the
magnetic structure of the sample is represented by a set of magnetic
variables $\{\theta_j,\varphi_j\}$ that describes orientation of FM
and AFM vectors inside the domains  ($j=$A1, A2, B1, B2) and a set
of variables $\{\xi_j\}$ that represents the amount of matter (say,
volume fraction) in the state of $j$-type (obviously, $\sum
\xi_j=1$). Equilibrium DS in presence of the external field is then
found from the condition of minimum of $\Phi$ with respect to
$\{\theta_j,\varphi_j, \xi_j\}$.

In such an approach one can neglect a contribution of the domain
walls into free energy potential $\Phi$. However, we implicitly
account for the inhomogeneities in space distribution of the FM and
AFM vectors when we chose independent variables for the potential
$\Phi$. Namely, reconstruction of the DS may proceed in two ways:
\emph{i}) by the field-induced motion of the domain walls;
\emph{ii}) by nucleation and growth of the energetically favorable
domains. The first way is almost activation-less while in the second
case the system should overcome the potential barrier related with
the formation of the domain walls. In the case under consideration
the domain walls between AFM (A/B) and FM (A1/A2 or B1/B2) domains
have different energies, and so, appear at different conditions. In
what follows we consider some typical situations and show the way to
control the DS with appropriate treatment of the sample.

\subsection{Four types of domains}
\begin{figure}[htbp]
 \includegraphics[width=0.5\columnwidth]{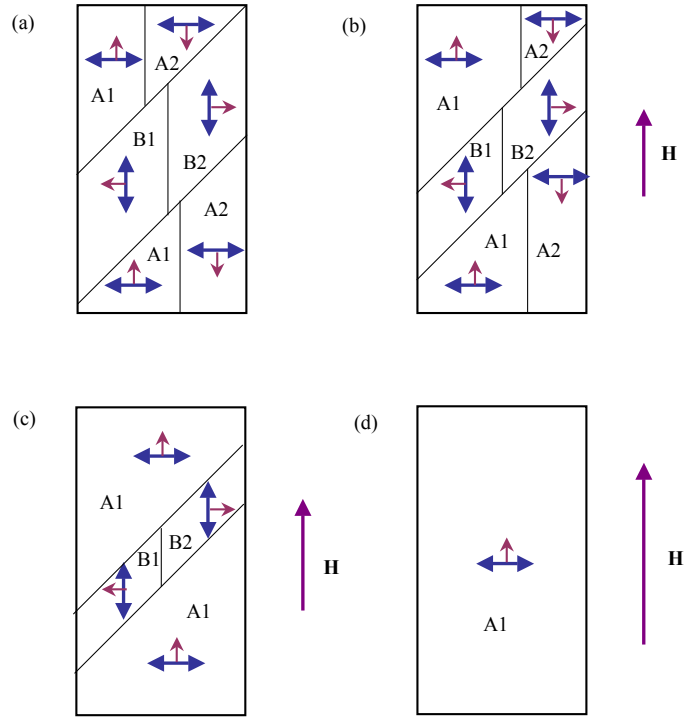}
  \caption{(Color online) Behavior of the combined FM and AFM domain structure in the external magnetic field
  $\mathbf{H}\|$[110] (parallel to the long side of the sample).
  (a) In the field absence all types of domains (A1, A2, B1, B2) are equally represented, the disbalance $\xi^{(0)}$ between A and B types depends upon the aspect ratio of the sample.
  (b) Below the
  first critical field, $H\le H_{\rm cr1}$, the domains are rearranged in such a way that the
   effective magnetic field vanishes.
  (c) In the interval $H_{\rm cr1}\le H\le H_{\rm cr2}$ the unfavourable domains of A2 type disappear, A1-domains compete with
  the domains of B-type.
  (d) Above the second critical field, $H >H_{\rm cr2}$, the sample is  a single domain.
  \label{fig_fm_plus_afm_H}}
\end{figure}

In the case when all four types of domains may freely grow or
diminish in size (say, in a virgin sample that initially contains
domains of all types), the external magnetic field is screened by an
appropriate domain configuration (see Fig.~\ref{fig_fm_plus_afm_H})
and the effective field inside the sample is zero. Equilibrium
values of the magnetic variables in this case are given by
Eq.~(\ref{equilibrium_domains}) and the domain fractions depend on
magnetic field as follows:
\begin{equation}\label{domain_fraction_4dom}
  \xi_{\rm A1,A2}=\frac{1}{4}\left[1-\xi^{(0)}\pm\frac{2H}{H_{\rm dm}}+\left(\frac{H}{H_{\rm
  des}}\right)^2\right];\quad\xi_{\rm B1}=\xi_{\rm B2}=\frac{1}{4}\left[1+\xi^{(0)}-\left(\frac{H}{H_{\rm
  des}}\right)^2\right],
\end{equation}
where we have introduced the demagnetizing, $H_{\rm dm}$, and the
destressing, $H_{\rm  des}$, fields:
\begin{equation}\label{demag_field-def}
    H_{\rm dm}\equiv
m_{\rm F}N^{\rm{dm}}_aM_0, \quad H_{\rm  des}\equiv 8\sqrt{J_0N^{\rm
des}}.
\end{equation}  
As seen from Eq.~(\ref{demag_field-def}), the value of destressing
field is enhanced due to exchange interactions (constant $J_0$). On
the contrary, the demagnetizing field is weakened due to small FM
moment ($m_{\rm F}\ll 1$). So, in the crystal under consideration
the demagnetizing effects are much smaller than the destressing
ones, $H_{\rm dm}\ll H_{\rm
  des}$ (see Table~\ref{table_calc}).

  The value $\xi^{(0)}$ introduced in
  Eq.~(\ref{domain_fraction_4dom}) represents the disbalance between type A and type B domains
in the absence of field. This value depends upon the shape of the
sample (or, equivalently, from the aspect ratio, see
Eq.~(\ref{destress_constants_thin_film})):
\begin{equation}\label{zero_field_fraction}
\xi^{(0)}\equiv \frac{N^{\rm des}_{\rm 2an}}{N^{\rm des}}\approx
\frac{c}{b}J_2(k).
\end{equation}
Such a shape-induced nonequivalence of domains has a magnetoelastic
origin (see Ref.\onlinecite{gomonay:174439} for details) and
originates from the AFM properties of the system. The disbalance
between type A and type B domains was noticed in the experiments
Ref.\onlinecite{Parks:2001} for the different sample shapes. The
value $\xi^{(0)}=0.22$ calculated from
Eq.(\ref{zero_field_fraction}) for the typical sample size (see
Table~\ref{table_calc}) fits well the experimental magnetization
curves, as we will see below.

The described configuration of the DS (see
Eq.~(\ref{domain_fraction_4dom})) is schematically shown in
Fig.~\ref{fig_fm_plus_afm_H}b. The fraction of the unfavourable
domains A2, B1 and B2 diminishes and at the critical value
\begin{equation}\label{first_critical}
  H=H_{\rm cr1}\equiv\frac{H^2_{\rm
  des}}{H_{\rm dm}}\left[1-\sqrt{1-(1-\xi^{(0)})\left(\frac{H_{\rm dm}}{H_{\rm
  des}}\right)^2}\right]\approx\frac{1}{2}(1-\xi^{(0)})H_{\rm dm}
\end{equation}
the unfavourable FM domains A2 disappears ($\xi_{\rm A2}=0$).

At $H\ge H_{\rm cr1}$ the internal effective magnetic field is
nonzero and magnetizations in the domains of B type rotate. However,
if $H_{\rm cr1}\ll H_{\rm s-f}$ (as, indeed the case in the crystal
under consideration), small tilt of $\mathbf{M}_{\rm F}$ and
$\mathbf{L}$ vectors can be neglected and field dependence of the
domain fractions (shown in Fig.~\ref{fig_fm_plus_afm_H}c) is
approximated as
\begin{equation}\label{domain_fraction_3dom}
  \xi_{\rm A1}=\frac{1}{2}\left[1-\xi^{(0)}+\frac{16Hm_{\rm F}J_0}{H^2_{\rm
  des}}+\left(\frac{H}{H_{\rm
  des}}\right)^2\right],\quad\xi_{\rm B1,2}=\frac{1}{4}\left[1+\xi^{(0)}-\frac{16Hm_{\rm F}J_0}{H^2_{\rm
  des}}-\left(\frac{H}{H_{\rm
  des}}\right)^2\right].
\end{equation}

The second critical field at which the unfavourable domains of B
type disappear ($\xi_{\rm B1,2}=0$) is given by the expression
\begin{equation}\label{second_crit_field}
H_{\rm cr2}\equiv 8m_{\rm F}J_0\left[\sqrt{1+\left(\frac{H_{\rm
des}}{8m_{\rm
F}J_0}\right)^2\left(1+\xi^{(0)}\right)}-1\right]\approx\frac{2N^{\rm
des}}{m_{\rm F}}\left(1+\xi^{(0)}\right).
\end{equation}


Above the second critical field, $H\ge H_{\rm cr2}$, the sample is a
single domain (A1) in average, with the possible remnants of the
states A2, B and corresponding domain walls that can serve as the
nucleation centers during the field cycling. Full monodomainization
of the sample takes place above the critical field $H_{\rm s-f2}\gg
H_{\rm cr2}$ at which all the states except A1 became unstable. 

Field cycling of the sample that initially had all the types of
domains is reversible if the maximal field value $H_{\rm max}$ is
not very large, $H_{\rm cr2}\leq H_{\rm max}\ll H_{\rm s-f2}$.
Macroscopic magnetization is parallel to the direction of the
external field due to the full compensation of the perpendicular
component by B1 and B2 domains.

\begin{table*}\label{table_calc}
\caption{Parameters used in numerical simulations. The source of
data (experimental or calculated) is specified in the last column.}
\begin{ruledtabular}
\begin{tabular}{lccc}
 Parameter & Meaning  & Value &
Rem      \\
 $a\times b\times c$& Sample
size&$7\times2\times0.5$~mm$^3$&Ref.\onlinecite{Parks:2001}\\ \hline
$m_F$& $M_F/M_0$&$7\cdot 10^{-4}$&Ref.\onlinecite{Kim:2001} \\
\hline $M_F$& Saturation magnetization &$7\cdot
10^{-3}$~emu/g&Ref.\onlinecite{Parks:2001}
\\ \hline $\xi^{(0)}$ & Shape-induced bias
& 0.22 & Eq.(\ref{zero_field_fraction})\\ \hline $H_{\rm dm}$ &
Demagnetization field & 0.3 Oe & Eq.(\ref{demag_field-def})\\ \hline
$N^{\rm des}$ & Destressing const., $T=120$K & $7$~mOe& Fitting \\
 &  $T=100$~K&1.5~mOe & param.\\ \hline
 $H_{\rm des}$ & Destressing
field, $T=120$~K&$2.1$~kOe&Eq.(\ref{demag_field-def})\\
 &
$T=100$~K&$1.1$~kOe&
\end{tabular}
\end{ruledtabular}
\end{table*}

\begin{figure}[htbp]
 \includegraphics[width=0.7\columnwidth]{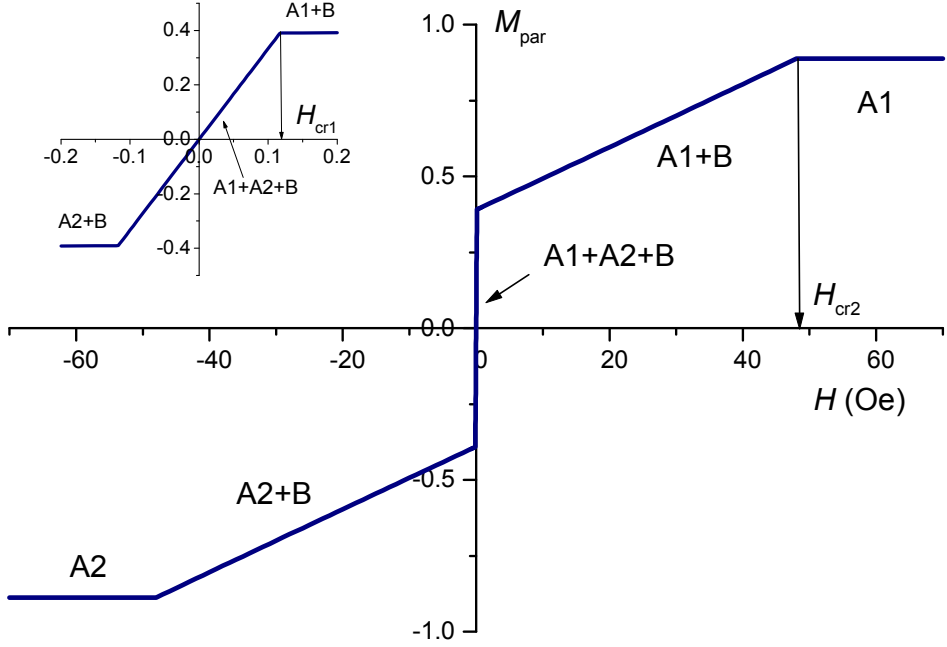}
  \caption{(Color online) Magnetization curve (projection on field direction) in the external magnetic field
  $\mathbf{H}\|$[110] for the case depicted in Fig.~\ref{fig_fm_plus_afm_H}.
  Inset shows the details of magnetization behavior below $H_{\rm cr1}$.
  The jogs (shown with arrows) arise at the critical fields $H=H_{\rm cr1,2}$
   when one type of domains disappear. Magnetization is normalized to saturation value.\label{fig_4dom_type_magnetization}}
\end{figure}

Field dependence of macroscopic magnetization $M_{\rm par}\propto
(\xi_{\rm A1}-\xi_{\rm A2})$ at $T=120$~K calculated from
Eqs.~(\ref{domain_fraction_4dom}) and (\ref{domain_fraction_3dom})
(see Tables \ref{table_data} and \ref{table_calc}) is represented in
Fig.~\ref{fig_4dom_type_magnetization}. One can distinguish three
intervals that correspond to different domain composition: \emph{i})
steep growth of $M_{\rm par}$ from 0 to $\propto
0.5(1-\xi^{(0)})M_{\rm F}$ (at $H=H_{\rm cr1}$) due to the motion of
A1/A2 domain walls initiated by demagnetization; \emph{ii}) smooth
growth of $M_{\rm par}$ from $\propto 0.5(1-\xi^{(0)})M_{\rm F}$  to
$\approx M_{\rm F}$ (at $H=H_{\rm cr2}$) due to the motion of A1/B
domain walls initiated by the destressing; \emph{iii}) very smooth
growth of $M_{\rm par}$ due to rotation of sublattice magnetizations
(not seen in Fig.~\ref{fig_4dom_type_magnetization}). Such a
behavior contrasts with a ``standard'' magnetization curve of FM and
also with the case when only two types of domains could compete
under the action of external field. The last case will be considered
in details in the next section.

\subsection{Competition of two domains}\label{two_domains}
Let us consider a sample that was preliminary monodomainized to the
state A1 by excursion into the region of high field, $H\geq H_{\rm
s-f2}$. The DS in this case depends upon the relation between
nucleation energies of different states.
As it was shown above, at $H>0$ an AFM domain B is more favorable
than a FM domain A2 (see  Fig.~\ref{fig_homog_vs_field}a). If, in
addition, there is a slight misalignment between the magnetic field
$\mathbf{H}$ and a crystal axis [110] that removes degeneracy
between the B1 and B2 states, the DS of the sample is represented by
the domains of only two types, A1 and B1.

Equilibrium values of the magnetic variables in this case were
calculated by the numerical minimization of the potential
(\ref{magnetic_energy_2}) with limitations $\xi_{A2}=\xi_{B2}=0$.
The values of the destressing coefficient $N^{\rm des}$ at different
temperatures (see Table \ref{table_calc}) were defined from the
fitting of experimental data \cite{Parks:2001}.

\begin{figure}[htbp]
  \includegraphics[width=0.7\columnwidth]{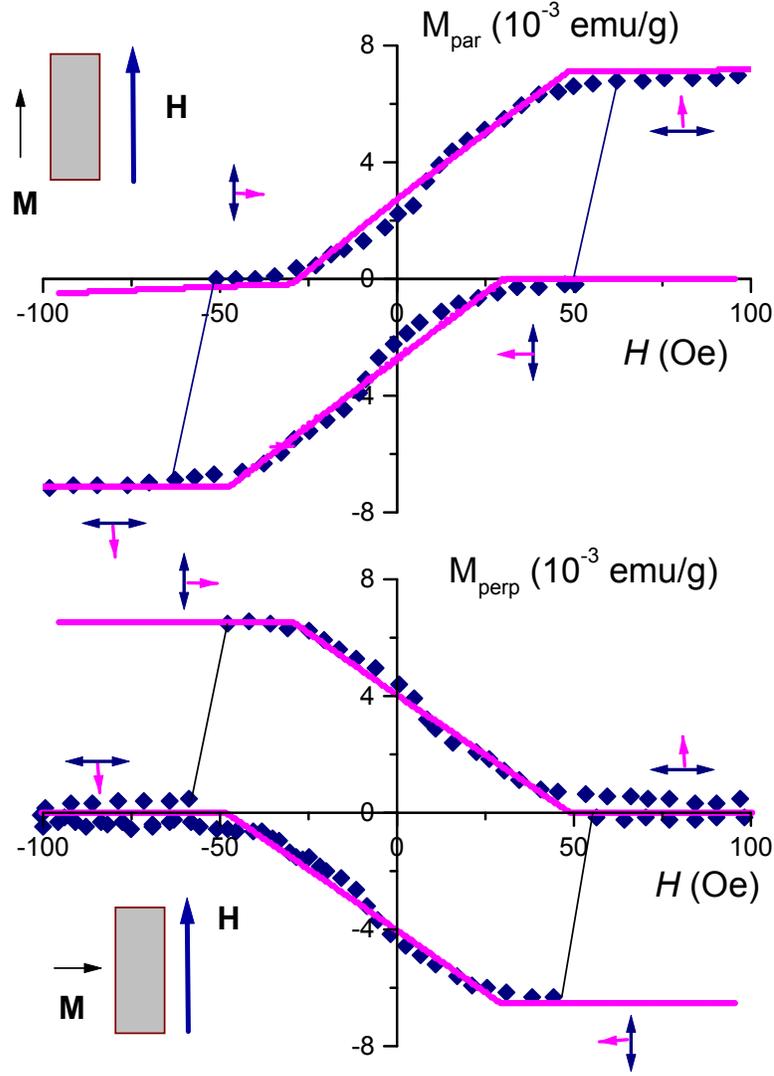}
  \caption{(Color online) Macroscopic magnetization \textit{vs} magnetic field.
  Points -- experimental data for Sr$_{2}$Cu$_{3}$O$_{4}$Cl$_{2}$ \cite{Parks:2001} taken at $T=$120~K after
  monodomainization of the sample at high fields $H\propto 5$~T.
  Solid lines -- theoretical approximation (see text for details).
  Upper and lower panels show, correspondingly,
  the parallel and perpendicular components of magnetization with respect to magnetic field.
  Insets show geometry of the experiment: orientation of the field with respect to the sample and orientation
  of the measured magnetization with respect to $\mathbf{H}$. The dominant type of domains for each field interval
  is depicted schematically by the single- and double-headed arrows. \label{fig_magnet_120}}
\end{figure}

Field dependence of macroscopic the magnetization at $T=120$~K
 is shown in Fig.~\ref{fig_magnet_120} with solid lines. Points represent experimental data \cite{Parks:2001}.
 Due to the fact that the domains A1 and B1
could not screen the external field, the macroscopic magnetization
has two components: one that is parallel to $\mathbf{H}$ (upper
panel in Fig.~\ref{fig_magnet_120}) and one that is perpendicular to
$\mathbf{H}$ (lower panel). The parallel and perpendicular
components represent the fractions of A1 and B1 domains,
respectively.

\begin{figure}[htbp]
  \includegraphics[width=0.7\columnwidth]{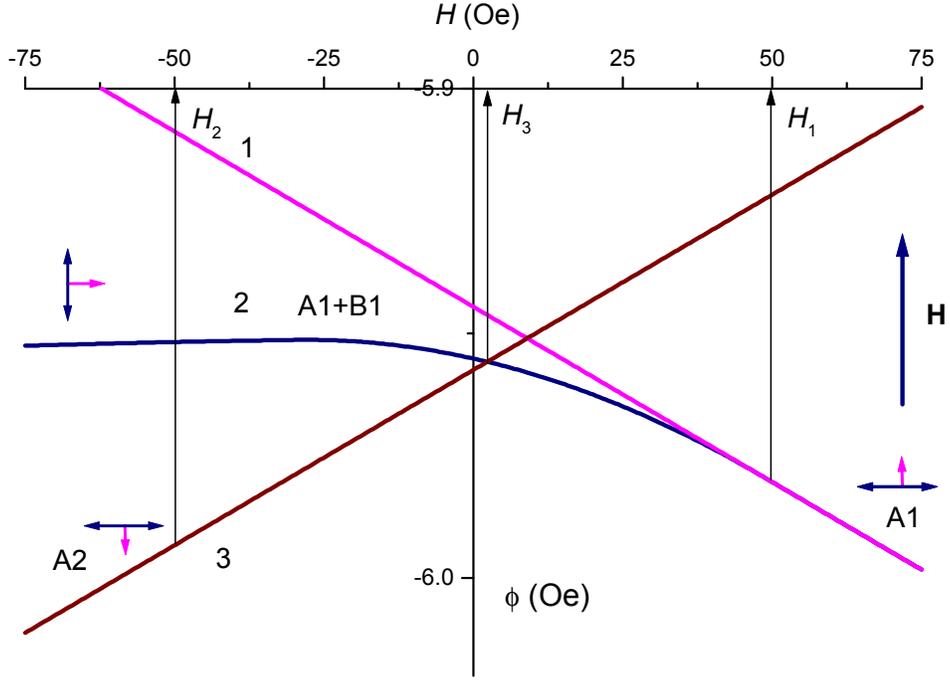}
  \caption{(Color online) Magnetic energy \textit{vs} magnetic field, $T=120$~K.
  Lines 1 and 3 correspond to a single domain state (domains A1 and A2, correspondingly),
  line 2 represents equilibrium two-domain state (domains A1 and B1).
Domain B1 appears at $H=H_1$. Domain A2 appears at field $H=H_2$
(determined empirically) when the energy difference between
two-domain and single domain states is enough for nucleation of this
energetically favorable domain. At $H=H_3$ the energy of two-domain
state is equal to the energy of a single-domain state A2, but the
potential barrier between two states prevents nucleation of the
domain A2. \label{fig_energy_120}}
\end{figure}

When the field decreases
  from high positive values, an AFM domains of B type appear and
magnetizations $M_{\rm par}(H)$ and $M_{\rm perp} (H)$ vary smoothly
between zero and saturation value. The slope of magnetization curves
depends upon the destressing coefficient and is thus much smaller
than the initial steep slope in the 4-domain case (see
Fig.~\ref{fig_4dom_type_magnetization}). At small negative field the
sample is almost a single domain (type B). However, this state is a
metastable one from the energy point of view, as seen from
Figs.\ref{fig_energy_120} and \ref{fig_homog_vs_field}a. Really,
below $H=H_3\approx 2.8$~Oe (marked with arrow in
Fig.\ref{fig_energy_120}) the energy of the state A2 (with
$\mathbf{M}_F\uparrow\downarrow\mathbf{H}$) is lower than that of a
single domain state B1 and a multidomain state A1+B1. On the other
hand, due to preliminary high-field treatment, the sample contains
no nucleation centers of A2 state. So,  the states B1 and A2 are
separated with the potential barrier that could be overcome only at
$H=H_2\approx -50$~Oe (according to Ref.\onlinecite{Parks:2001},
this value varies from sample to sample and depends on temperature).
After the subsequent excursion into high negative fields (well below
$H_{2}$) the sample transforms into a single domain A2 and one can
observe competition between the A2 and B2 domains during the further
field increase.
\subsection{Domain structure and field treatment}
In the previous subsections we have considered two limiting cases of
field treatment that result in two types of magnetization curves. In
the virgin sample (no field treatment) the magnetization can be
smoothly and reversibly changed between two opposite directions.
Field cycling between high fields (high enough to remove all the
domain walls and the remains of unfavorable domains) results in a
hysteretic behavior when magnetization varies smoothly between zero
and saturation value and then suddenly changes due to transition
from metastable to stable state.

\begin{figure}[htbp]
  \includegraphics[width=0.7\columnwidth]{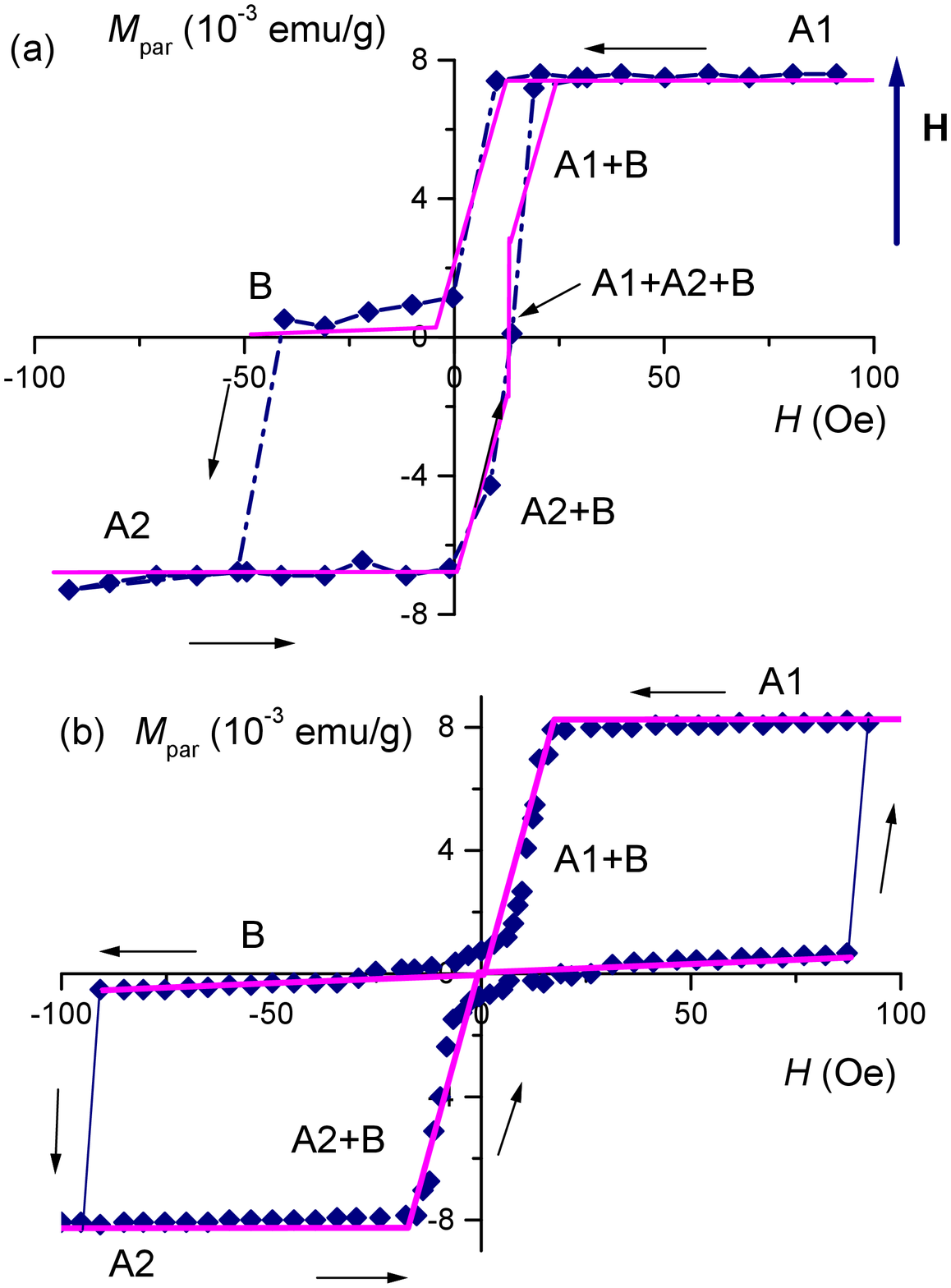}
  \caption{(Color online) Macroscopic magnetization \textit{vs} magnetic field for different field treatment.
   Points -- experimental data for Sr$_{2}$Cu$_{3}$O$_{4}$Cl$_{2}$ \cite{Parks:2001} taken at $T=$100~K (see text for
  details), solid lines -- theoretical
approximations. Thin arrows show the direction of field sweeping.
(a) Competition between FM and AFM domain structure. Field treatment
starts from the high positive values ($H\approx 1$~T) where the
sample is a single domain. Variation of field is swept at
intermediate value $|H_2|\le|H|\ll|H_{\rm cr2}|$, large enough to
induce switching between metastable, B1, and stable, A2, states and
small enough to remove the traces of A1 and B1 phases from the
sample. (b) Hysteresis loop with the excursion into high fields. For
any field value the DS includes only two types of domains, as in
Fig.~\ref{fig_magnet_120}. Only parallel component of magnetization
is shown. \label{fig_magnet_100}}
\end{figure}

In this subsection we consider some intermediate case when a single
domain sample is cycled in low fields. Corresponding magnetization
curve is shown in Fig.~\ref{fig_magnet_100}a (solid lines --
numerical simulations, points -- experimental data
\cite{Parks:2001}, $T=100$~K). Field cycling starts at high positive
fields, where the sample is a single domain. When the field is
decreased down to $H=H_{1}$ (see Fig.~\ref{fig_energy_120}a) the
domains of B1 type appear and the DS consists of A1 and B1 domains.
Magnetization curve (upper curve in Fig.~\ref{fig_magnet_100}a)  in
this case is of two-domain type discussed in Subsection
\ref{two_domains} (we still assume slight misalignment that excludes
one type of B domains). However, further behavior of the DS and
hence, magnetization, depends upon the size of the loop. If the loop
is small ($|H|\le |H_2|$, where $H_2$ is a coercive field at which
domain B1 transforms into A2 as explained above), magnetization
varies smoothly between zero value at negative fields and saturation
value at positive fields. If the loop is very large ($|H|\ge |H_{\rm
cr2}|$), the DS structure consists of two domains: A1 and B1 for
large positive and small negative fields and A2 and B2 for large
negative and small positive fields, as shown in
Figs.~\ref{fig_magnet_100}b and \ref{fig_magnet_120}. In the
intermediate case ($|H_2|\le|H|\ll|H_{\rm cr2}|$) the DS includes
three types of domains, A1, A2 and B2 (lower curve in
Fig.~\ref{fig_magnet_100}a), and magnetization curve is asymmetric.
It is worth to note that theoretical  magnetization curves
calculated with only one fitting parameter (destressing coefficient
$N^{\rm des}$) fit well experimental data, as seen from
Figs.\ref{fig_magnet_100} and \ref{fig_magnet_120}. 

\section{Discussion}
We have considered the different types of the DS behavior in a
multiferroic with AFM and FM order parameters. Depending on the
field treatment the DS may include from one to four types of domains
and can be unambiguously determined from the magnetization curves in
the field $\mathbf{H}\|[110]$. Namely, if DS includes all types of
domains, macroscopic magnetization is parallel to $\mathbf{H}$,
magnetization curve is reversible, varies between $\pm$ saturation
value, and includes steep section at small fields. If the DS
includes three types of domains A1, A2 and B1, the macroscopic
magnetization has two components, parallel, $M_{\rm par}$, and
perpendicular, $M_{\rm perp}$, to $\mathbf{H}$. During field cycling
$M_{\rm par}$ varies between positive and negative saturation
values, while $M_{\rm perp}$ varies between zero and saturation
value. At last, if the DS includes only two domains, A and B, both
$M_{\rm par}$  and $M_{\rm perp}$ vary between zero and saturation
value.

We argue that formation of AFM (B) domain results from the
destressing effect which, in turn, originates from magnetoelastic
interactions. An absolute value of magnetoelastic constant is rather
small (compared to such AFMs as NiO, KCoF$_3$, etc) and corresponds
to spontaneous strain $u\propto 10^{-6}$ (for estimation we took
$c_{44}=20$~GPa at $T=120$~K). Such a small value of $u$ explains
the low potential barrier for formation of AFM domains.

\begin{figure}[htbp]
  \includegraphics[width=0.7\columnwidth]{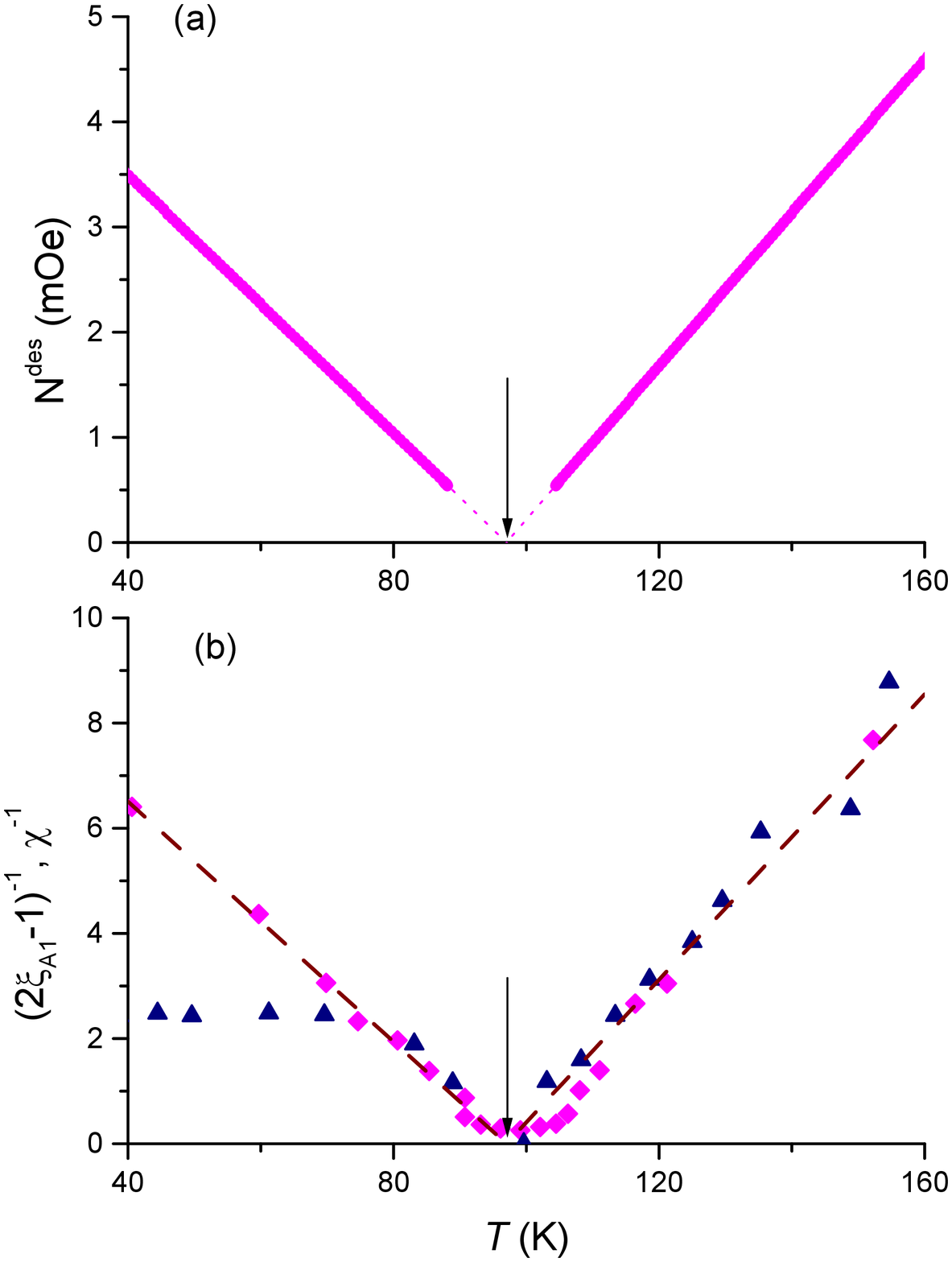}
  \caption{(Color online) Temperature dependence (a) of the
  destressing coefficient $N^{\rm des}$ predicted from the comparison with the
  (b) temperature dependence of the
  reciprocal domain A1 fraction $(2\xi_{\rm A1}-1)^{-1}$ (triangles) and the inverse
  susceptibility $\chi^{-1}$ (diamonds) plotted according to data
  \cite{Parks:2001}. Raw data for $\chi^{-1}$ were normalized (multiplied by appropriate factor) to fall into the same ranges of values as $(2\xi_{\rm A1}-1)^{-1}$. Dash line shows linear
  approximation of the experimental data. Peculiarity at $T=97$~K is indicated with arrow. \label{fig_temper_depend}}
\end{figure}

Analysis of magnetization curves shows that low-field susceptibility
$\chi$ of the sample that consists of AFM domains is inversely
proportional to the destressing coefficient $N^{\rm des}$ (in
contrast to FM, where $\chi$ depends upon demagnetization constant).
According to the experiments \cite{Parks:2001}, the inverse
susceptibility $\chi^{-1}$ of Sr$_{2}$Cu$_{3}$O$_{4}$Cl$_{2}$ shows
nontrivial temperature dependence (see Fig.\ref{fig_temper_depend}b)
and attains the minimum at $T=T_0=97$~K. The domain fraction
$\xi_{\rm A1}$ at fixed $\mathbf{H}$ extracted from the neutron
scattering experiments of the reminded group \cite{Parks:2001} shows
the same temperature dependence as $\chi$, as can be seen from
Fig.\ref{fig_temper_depend}b. Using correlation between $N^{\rm
des}$ and $\chi$ we predict the following temperature dependence of
the destressing coefficient depicted in
Fig.\ref{fig_temper_depend}a:
\begin{equation}\label{temperature_dep_destres}
 N^{\rm
des}(T)=\left\{\begin{array}{cc}
    7.3\cdot 10^{-5}\cdot(T-T_0),& T\ge T_0, \\
    6.13\cdot 10^{-5}\cdot(T_0-T),& T<T_0.
 \end{array}
 \right.
\end{equation}

If we take into account that $N^{\rm des}\propto \lambda^2/c_{44}$
(see Eq.(\ref{destress_constants_thin_film})) we may also anticipate
a peculiarity of the elastic (or magnetoelastic) properties of the
crystal in the vicinity of $T=T_0$.

In summary, we described the possible scenario of field-induced
restructurization of the domains in the system that consists of the
domains of different physical nature. The proposed model can be
extended to multiferroics that show simultaneously ferroelectric and
AFM ordering and also to FM martensites with ferroelastc and
ferrpmagnetic ordering.

\begin{acknowledgements}
The authors acknowledge the financial support from the Department of
Physics and Astronomy of he National Academy of Sciences of Ukraine
in the framework of Special Programme for Fundamental Research. The
work was partially supported by the grant from the Ministry of
Education and Science of Ukraine.
\end{acknowledgements}
%

\end{document}